\documentclass[conference]{IEEEtran}
\IEEEoverridecommandlockouts
\usepackage{cite}
\usepackage{amsmath,amssymb,amsfonts,bbm}
\usepackage{graphicx}
\usepackage{textcomp}
\usepackage{booktabs}
\usepackage{xcolor}
\usepackage{bm}
\usepackage{subcaption}
\usepackage[numbers]{natbib}

\usepackage[utf8]{inputenc}
\usepackage[ruled,vlined]{algorithm2e}
\DontPrintSemicolon
\usepackage{algpseudocode}
\usepackage{multirow}
\usepackage{setspace}
\usepackage{array}
\newcolumntype{P}[1]{>{\centering\arraybackslash}p{#1}}
\def\BibTeX{{\rm B\kern-.05em{\sc i\kern-.025em b}\kern-.08em
    T\kern-.1667em\lower.7ex\hbox{E}\kern-.125emX}}

\begin{document}

\title{Communication Load Balancing via Efficient Inverse Reinforcement Learning}

\author{\IEEEauthorblockN{Abhisek Konar$^*$, Di Wu$^*$, Yi Tian Xu$^*$, Seowoo Jang$^\dagger$, Steve Liu$^*$, Gregory Dudek$^*$}
\IEEEauthorblockA{$^*$Samsung Electronics, Canada,  $^\dagger$Samsung Electronics, Korea (South)\\
\{abhisek.k, di.wu1, seowoo.jang, steve.liu, greg.dudek\}@samsung.com}
}

\maketitle

\begin{abstract}


Communication load balancing aims to balance the load between different available resources, and thus improve the quality of service for network systems. After formulating the load balancing (LB) as a Markov decision process problem, reinforcement learning (RL) has recently proven effective in addressing the LB problem. To leverage the benefits of classical RL for load balancing, however, we need an explicit reward definition. Engineering this reward function is challenging, because it involves the need for expert knowledge and there lacks a general consensus on the form of an \textit{optimal} reward function. In this work, we tackle the communication load balancing problem from an inverse reinforcement learning (IRL) approach. To the best of our knowledge, this is the first time IRL has been successfully applied in the field of communication load balancing. Specifically, first, we infer a reward function from a set of demonstrations, and then learn a reinforcement learning load balancing policy with the inferred reward function. Compared to classical RL-based solution, the proposed solution can be more general and more suitable for real-world scenarios.  Experimental evaluations implemented on different simulated traffic scenarios have shown our method to be effective and better than other baselines by a considerable margin.

\end{abstract}

\section{Introduction}

The volume of wireless communication data has been snowballing in recent years. As reported in~\cite{yahoo}, the annual mobile data has increased seven times in the past six years and it is expected to reach 220.8 million terabytes per month by 2026. 
Alongside surge in mobile traffic, their distribution is also very uneven.  As reported in ~\cite{holma2012LTE-Advanced}, over $50\%$ of the mobile traffic is being routed through a number as low as $15\%$ of the existing cells, severely hampering the quality of experience of the users being served by these overloaded cells. Load balancing aims to balance the traffic distribution within the network, improving the quality of service (QoS) of the systems and the quality of experience (QoE) for the customers. 



Load balancing algorithms can be classified into two broad categories: rule-based methods, and reinforcement learning-based methods. Rule-based methods aim to balance the load distribution using pre-determined rules but they usually lack the ability to adapt to rapidly evolving network conditons. Reinforcement learning (RL) is a class of learning algorithms, where a controller is optimized by interacting with an environment. RL has recently shown impressive performance on communication load balancing \cite{yuexu2019drlMLB, wu2021dataefficientRL, feriani2022multiobjective, ma2022coordinated,kang2021hierarchical}. 
It is particularly suited for solving intricate tasks with well-defined rewards like Atari games \cite{mnih2013playing}. It has also been applied for real-world tasks such as self-driving cars \cite{sallab2017deep}, smart grid~\cite{wu2018optimizing,, fu2022reinforcement,zhang2022metaems,wu2018machine}, traffic control~\cite{huang2021modellight}. 
The outcome of an RL policy is contingent upon the design choices of the associated reward~\cite{fucloser}. Reward-engineering, i.e., the design of the RL reward function, for real-world tasks can be quite challenging. Especially, for tasks like load balancing in communication networks, where improvement in the QoE of the customers is gaining significant traction. But QoE can be vague and different network performance indicators can contribute to varying degrees to form the desired reward function that can effectively capture the QoE for a demographic. Inverse reinforcement learning (IRL)~\cite{abbeel2004irl} can act as a potential solution by circumventing the need for tedious reward designing instead inferring it from expert data.

IRL leverages a set of expert demonstrations to infer the underlying reward function that best explains the behavior of the expert.
It has been used for real-world problems that lack a well-defined reward function like in robotics \cite{finn2016guided, zhang2021learning} and autonomous driving \cite{zou2018inverse}. 
\begin{figure}[htp]
    \centering
    \includegraphics[width=0.85\columnwidth]{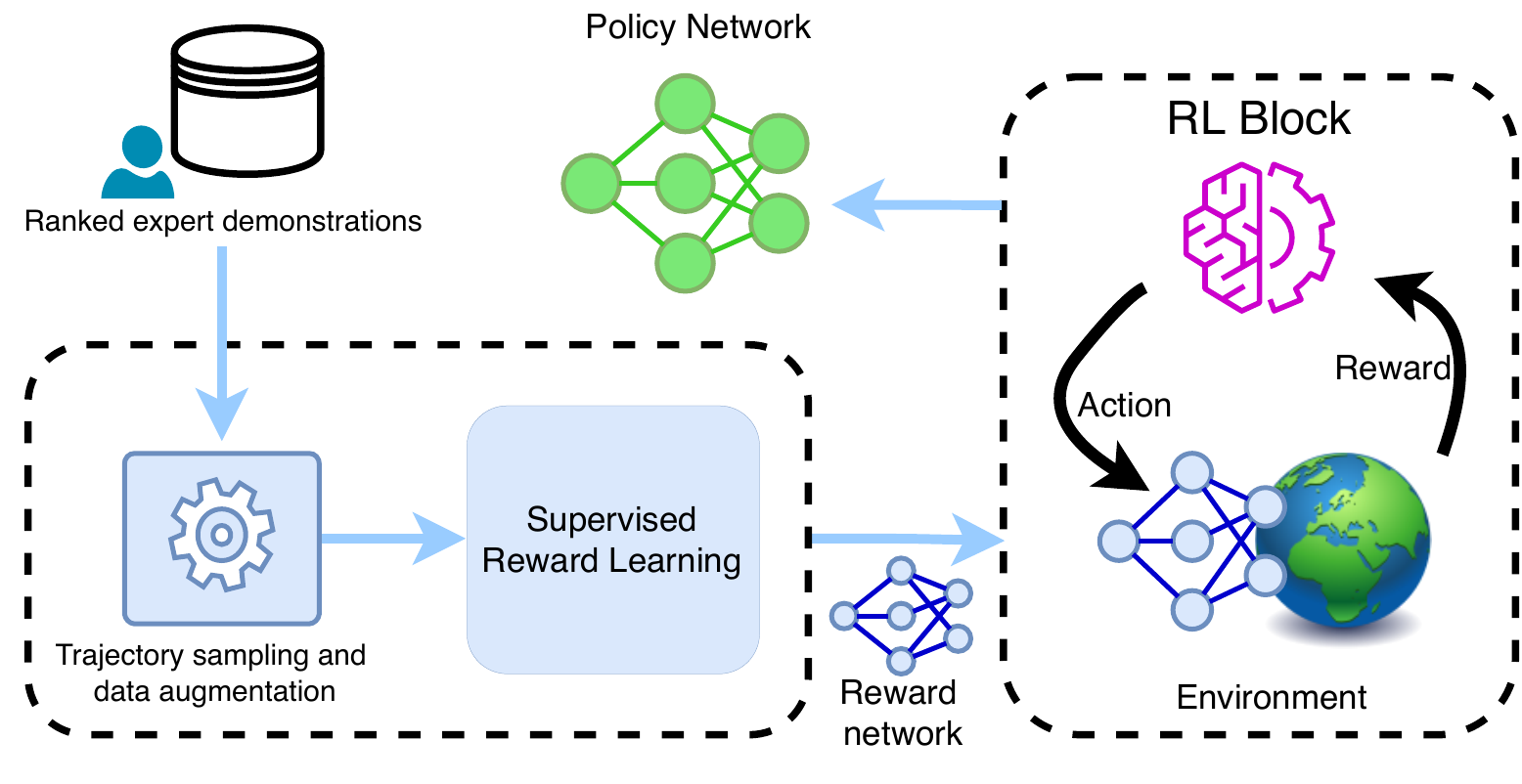}
    \caption{Overview of our IRL-based pipeline for load balancing. }
    \label{fig:tcs-trex-overview}
\end{figure}
Traditionally, IRL algorithms are generally 
formulated under the assumption that ``expert'' demonstrations are optimal and, the goal of such algorithms is to recover the reward function that best explains the observed expert behavior. This assumption, however, limits the performance of the output policy to that of the expert demonstrations, which in reality are rarely actually optimal. This limitation is addressed in a recent work on reward extrapolation~\cite{brown2019extrapolating}. Reward extrapolation is able is exploit sub-optimal demonstrations to infer a reward function which is used to train policies that can outperform the demonstrator. 

In this paper, we present, the first attempt of using IRL and reward extrapolation for communication load balancing. The contributions of this paper are two-fold: (1) an IRL-based learning framework (see Fig.\ref{fig:tcs-trex-overview}) to train a reward function that accurately captures the latent reward function from a set of ranked sub-optimal demonstrations. This reward function is used in a downstream RL task to train a policy function that significantly outperforms the demonstrations; and (2) a trajectory sub-sampling technique, Temporally Consistent Sampling (TCS), that is suited for load balancing.

\section{Background}
\subsection{Terminology and notation}
\begin{figure}[htbp]
    \centering
    \includegraphics[width=1\columnwidth]{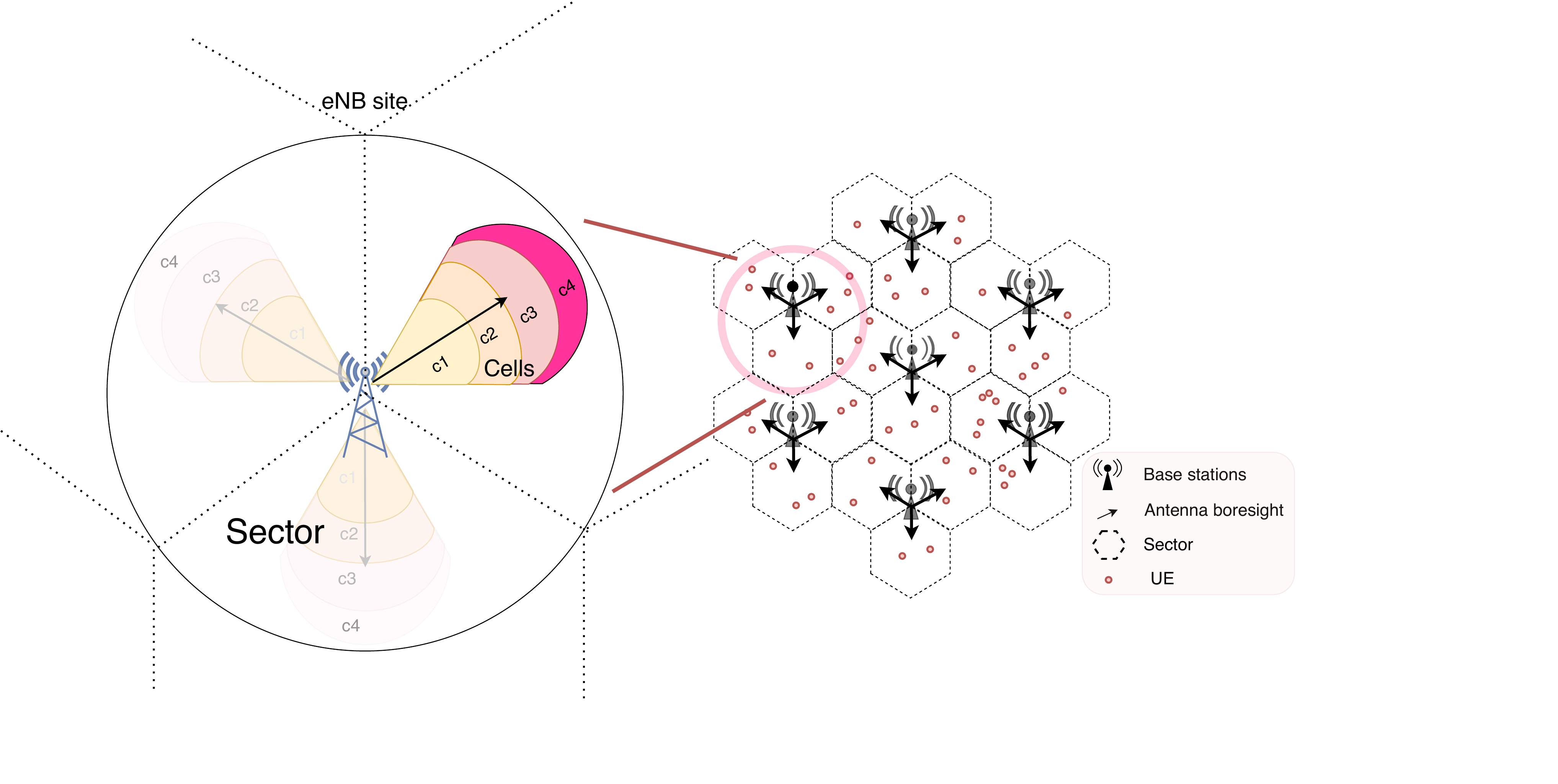}
    \caption{The simulated communication network layout with seven eNBs along with a graphical representation of a base station within the network.}
    \label{fig:hex7-layout}
\end{figure}

For consistency, we define a few terms that will be used throughout this paper as well as some mild simplifying assumptions. A communication network is composed of a number of base stations (eNB). An eNB is a physical site containing radio access devices. Each eNB in turn consists of $N_S$ sectors. Each sector is made up of $N_C$ cells ($c_1, c_2, ... c_{N_{c}}$), one for each frequency in the sector. A cell serves user equipment (UEs) of a particular carrier and is directional in nature. The sectors of an eNB are designed in a non-geographically overlapping fashion to maximize coverage. A schematic diagram of a base station is shown in Fig.\ref{fig:hex7-layout}.  In practice, although actual networks may violate some of these assumptions, they are needed only for explanatory purposes.

\subsection{Load balancing mechanisms}
Out of the many network load balancing mechanisms presently in practice, in this work, we focus on idle mode user equipment based load balancing and mobility load balancing. They differ by the status of the UEs that they redistribute, targeting idle and active UEs respectively.

\subsubsection*{Idle mode UE based Load Balancing}
IULB deals with offloading idle UEs from an overloaded cell to its neighbors by adjusting cell re-selection priorities. Each cell $i$ has an associated IULB weight ($w_i$). Once the load of a given cell breaches a specific threshold, idle UEs from that cell move to nearby cells with probability proportional to their current IULB weight. As IULB deals with idle UEs, improvements are perceived when the UEs becomes active.

\subsubsection*{Mobility Load Balancing}
MLB focuses on migrating active UEs from an overloaded cell to their less-loaded counterparts through handovers. 
MLB gets triggered when the IP throughput of a given cell (source cell) falls below a predetermined threshold value. Then, the source cell determines the set of UEs to transfer and their corresponding target cells according to the signal quality from the UEs and the IP throughput of the neighbouring cells. 
Unlike IULB, MLB has immediate effect on the network.

\subsection{Metrics for load balancing}
\label{subsection:kpis}
Many metrics, focusing on different aspects of the network, can be used to quantify the condition of a network. In this work, we use the IRL approach that, in theory, should bypass the need for reward engineering in favour of pairwise trajectory ranking provided by an expert. However, due to the lack of such infrastructure, we resort to using an engineered reward function. We focus on a set of metrics based on the throughput of the network following the practice used in a few recent works~\cite{wu2021dataefficientRL,kangHierarchy2021,Li2022clusteringRL}. 
\begin{enumerate}
    \item $\mathbf{T}_{min}$ = $\min_{i \in \{1,...N_{c}\}} x_i$ denotes the minimum IP throughput among the cells of the sector under consideration. Here, $x_i$ denotes the IP throughput of cell $i$. Higher $\mathbf{T}_{min}$ indicates better worst case cell performance.
    \item $\mathbf{T}_{std}$ = $\sqrt{\frac{1}{N_{c}} \sum^{N_c}_{i=1}(x_i- \frac{1}{N_c}\sum^{N_c}_{i=1}x_i})^2$ is the standard deviation in IP throughput of the cells. 
    Lower $\mathbf{T}_{std}$ implies fairer services across the cells. 
    \item $\mathbf{T}_{cc}$ = $\sum_{i=1}^{N_{c}} \mathbbm{1}(x_i < \mathbbm{x})$ counts the number of cells in the sector that have a throughput lower than a given threshold value $\mathbbm{x}$. Here $\mathbbm{x}$ is chosen to be a small constant. Lower $\mathbf{T}_{cc}$ suggests less congested cells.
\end{enumerate}
In addition to gauging performance, the above KPIs individually or as a combination can also serve other roles like creation of a reward function in an RL setting \cite{Li2022clusteringRL}.

\subsection{Load balancing algorithms}
Classic load balancing algorithms are rule-based and use expert domain knowledge to make informed decisions. \cite{kwan2010} performs intra-frequency load balancing in Long Term Evolution (LTE) networks by automatically adjusting the cell-specific offset based on the current cell loads. \cite{yingyang2012} focuses on the use of adaptive step-size to adjust the offset between neighboring cells. 
\cite{elbv_shorabi2021} exploits the geometric properties of the network infrastructure and uses Voronoi tessellations over a geographical area to compute UE association with nearby edge servers.
Data-driven approaches have recently enjoyed considerable success. Various learning-based methods, including supervised and reinforcement learning, have been successfully applied to the problem of load balancing.
\cite{zhang2021STD} use spatio-temporal network data to predict future traffic patterns and subsequently use this information to perform proactive user association.
\cite{mwanjeSON2016}and \cite{kudoHnets2014} use Q-learning to address the load balancing in self-organizing networks, and heterogeneous networks respectively. 
\cite{yuexu2019drlMLB} proposes a multi-agent actor-critic network model in a model-free off-policy setting to obtain an optimal policy for MLB.
\cite{gupta2021DRL} uses RNN to understand past SINR measurements as a function of UE trajectory and number of HOs. \cite{wu2021dataefficientRL} propose a data-efficient transfer learning-based RL approach that is robust to environmental fluctuations.

A thorough scrutiny, however, will reveal a lack in consensus for the choice of reward function. \cite{mwanjeSON2016} considers the change in load while optimizing for MLB. 
\cite{yuexu2019drlMLB} use the inverse of the maximum load of a cell in a given neighborhood and, \cite{kudoHnets2014} relies on the change in overload among neighboring cells. The choice of the reward function, in RL, defines the task \cite{Ng00algorithmsfor} and plays a pivotal role in determining the final performance of the controller policy. There are different opinions in the community on what an \textit{ideal} reward function should be. 

\subsection{Inverse RL and reward extrapolation}
Inverse RL aims to recover a reward function from expert demonstrations that best explains the expert's behavior \cite{abbeel2004irl}. 
While the use of IRL is yet to receive traction in network load balancing community, there has been recent work that use it in the field of cellular networks. \cite{zhang2022} use IRL to optimally allocate power in multi-user cellular networks. Other real-world applications where IRL has seen significant success are autonomous driving and dexterous robotic manipulation. In autonomous driving, expert demonstrations are used to train a policy that drives  \cite{zou2018inverse} and parks \cite{fang2021maximum} like a human driver.
For robotic manipulation, \cite{finn2016guided} uses IRL to train a robot to perform dexterous tasks. \cite{xie2019learning} trains a model to learn grasping from failed demonstrations. Recent works like TRajectory EXtrapolation (T-REX) \cite{brown2019extrapolating} aim towards reward extrapolation. The goal being able to leverage the goodness of expert data to come up with a reward function that can infer rewards from unseen states.  It is a relatively new topic, with applications limited to environments like Atari and Mujoco \cite{mujoco}.

\section{Proposed Method}
Our proposed method, TRajectory EXtrapolation(T-REX) \cite{brown2019extrapolating} using Temporally Consistent Sampling (TCS) follows a similar training pipeline of T-REX along with the inclusion of a data-augmentation module, that uses TCS, to generate training data for better extrapolation results. We start by defining the problem in Section \ref{subsection:problem_formulation}, followed by the individual components of the IRL-based learning framework in Sections \ref{subsection:reward_learning} and \ref{subsection:data_augmentation}.

\subsection{Problem formulation}
\label{subsection:problem_formulation}
The problem is modelled as an Markov decision process (MDP) represented by a 4-tuple ($\mathcal{S}$, $\mathcal{A}$, $\mathcal{P}$, $\mathcal{R}$) and are defined as follows:
\begin{enumerate}
    \item $\mathcal{S}$ is the set of states. Each state, s $\in \mathcal{R}^{3 \text{x}N_C}$, consists of 3 components: the number of active UEs $s_{ue} \in \mathcal{R}$, the average throughput per cell $s_{ip} \in \mathcal{R}$, and the average percentage of the physical resources of each cell used $s_{prb} \in \mathcal{R}$ of every cells of a given sector. The number of cells in our system, $N_C$, is 4. Hence s $\in \mathcal{R}^{12}$.
    \item $\mathcal{A}$ is set of actions. Each action, $a$ consists of two parts, one to initiate IULB ($a_{IULB} \in \mathcal{R}^{N_C}$) and one that controls handover thresholds to trigger MLB ($a_{MLB} \in \mathcal{R}^{3 \text{x} N_C}$). All the actions are discrete with a step size of $1$.
    \item $\mathcal{P}(s,a,s')=P(s'=s_{t+1}|s=s_t, a=a_t)$  are the state transition probabilities. 
    \item $\mathcal{R} : \mathcal{S} \rightarrow \mathbb{R}$ is the reward function.
\end{enumerate}
For IRL, $\mathcal{R}$ is unavailable and the objective is to learn the reward function from expert observations. A policy $\pi_{\phi} : \pi_{\phi}(a|s) \in [0, 1]$, parameterized by $\phi$, is a function that maps a given state, $s \in \mathcal{S}$, to a distribution over actions, $a \in \mathcal{A}$. A trajectory $\tau$ of length $n$ is represented by a sequence of states ${\{s_1, s_2, ... s_n\}}$.

Given a set of $m$ ranked trajectories, ${\{\tau_1, \tau_2, ... \tau_m\}}$ where  $\tau_j$ is better than $\tau_i$ (i.e., $\tau_i \prec \tau_j$) if $ i < j$, the objective is to find a parameterized reward network $\hat{r}_\theta$ that is able to capture the relative ranking of the demonstrated trajectories. In the process, it has to extrapolate the underlying reward function the demonstrations are trying to maximize. Once, $\hat{r}_\theta$ is obtained, it is used to train a policy that has the potential to outperform the demonstrations.

\subsection{Reward extrapolation using T-REX}
\label{subsection:reward_learning}
The goal is to train a reward network, $\hat{r}_{\theta}$, that maintains $\hat{J}_\theta(\tau_i) < \hat{J}_\theta(\tau_j)$ when $\tau_i \prec \tau_j$, where $\hat{J}_\theta(\tau_i) = \sum_{s \in \tau_i} \hat{r}_\theta(s)$ denote the total reward obtained by trajectory $\tau_i$ using $\hat{r}_\theta$. For such a model, the general loss function can be given by Equation \ref{eqn:trex-loss-general}.
\begin{equation}
    \label{eqn:trex-loss-general}
    \mathcal{L}(\theta) = \mathbf{E}_{\tau_i, \tau_j \in \Pi}\bigg{[} \xi(P(\hat{J}_\theta(\tau_i) < \hat{J}_\theta(\tau_j)), \tau_i \prec \tau_j \bigg{]}   
\end{equation}
where $\Pi$ is the set of ranked demonstrations, and $\xi$ is a binary classification loss. Following the classic models of preference \cite{bradley1952rank} (Equation \ref{eqn:preferencemodel}), and using a cross-entropy loss for $\xi$, $\mathcal{L}$ can be rewritten as Equation \ref{eqn:trex-loss-fn}.
 
\begin{equation}
    \label{eqn:preferencemodel}
    P\bigg{(}\hat{J}_\theta(\tau_i) < \hat{J}_\theta(\tau_j)\bigg{)} \approx \frac{\exp{\sum\limits_{s \in \tau_j} \hat{r}_\theta(s)} } { \exp{\sum\limits_{s \in \tau_i} \hat{r}_\theta(s)} + \exp{\sum\limits_{s \in \tau_j} \hat{r}_\theta(s)}}
\end{equation}
\begin{equation}
    \label{eqn:trex-loss-fn}
    \mathcal{L}(\theta) = - \sum_{\tau_i \prec \tau_j} \log \bigg{(} \frac{\exp{\sum\limits_{s \in \tau_j} \hat{r}_\theta(s)}} {\exp{\sum\limits_{s \in \tau_i} \hat{r}_\theta(s)} + \exp{\sum\limits_{s \in \tau_j} \hat{r}_\theta(s)} } \bigg{)}
\end{equation}

Once a reward network $\hat{r}_\theta$ is trained, it is used to train a policy using Proximal Policy Optimization (PPO) \cite{schulman2017proximal}. Additional details about the training parameters for both reward and policy learning are provided in Section \ref{subsection:training_details}.

\subsection{Temporally Consistent Sampling}
\label{subsection:data_augmentation}
\begin{figure}
    \centering
    \includegraphics[width=0.85\columnwidth]{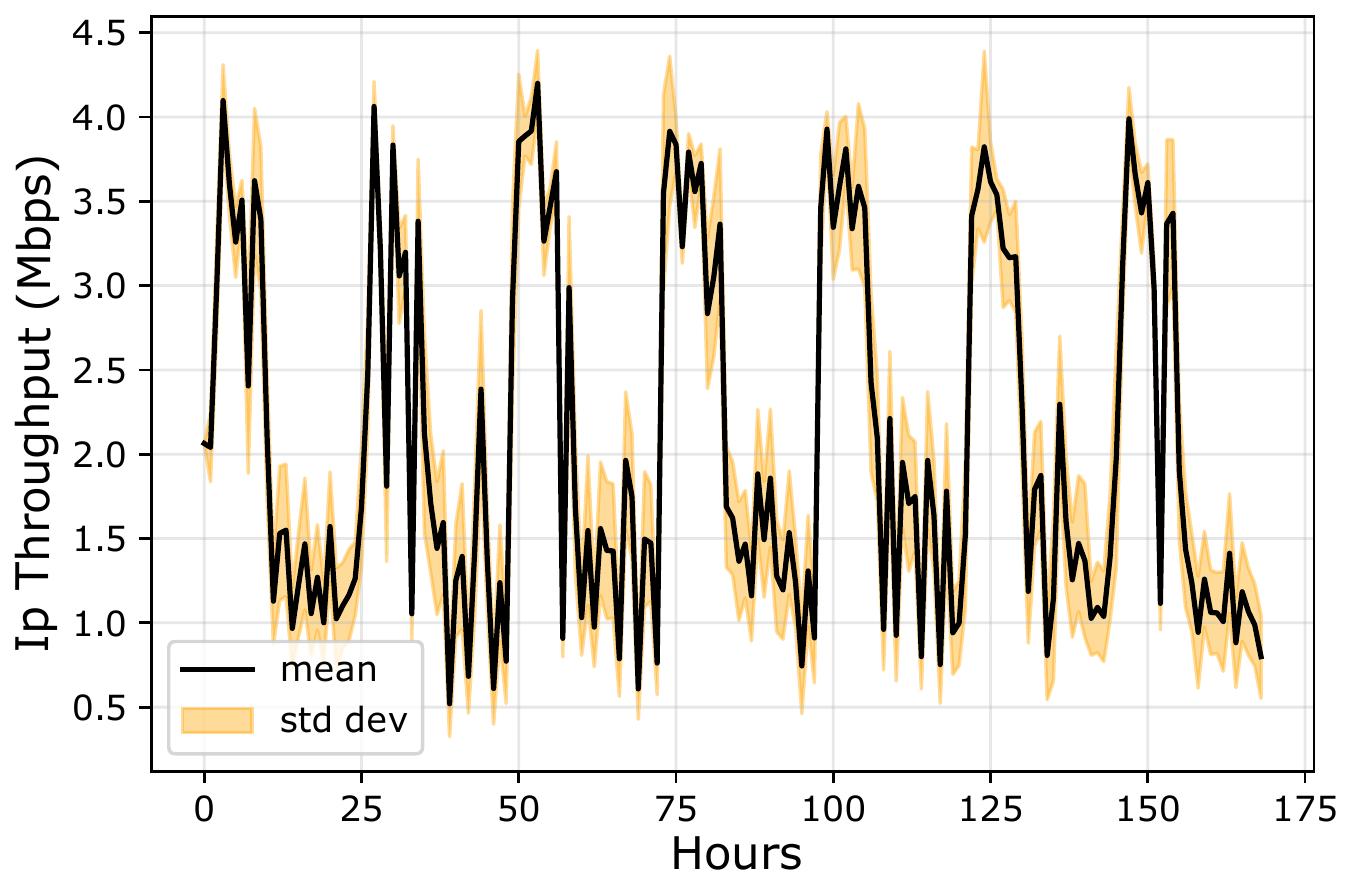}
    \caption{Plot of all trajectories from the demonstration set from a load balancing scenario. The trajectories exhibit a periodic behavior over the minimum IP throughput. Similar trends are observed across different network KPIs.}
    \label{fig:trajectory_over_timesteps}
\end{figure}

The collection of expert demonstrations is expensive, and reward extrapolation relies on data augmentation making the sub-sampling technique employed for the data augmentation a vital component of the training pipeline. \cite{brown2019extrapolating} proposes the sampling of contiguous blocks of states of equal length from randomly selected starting points in trajectories. The method assumes that the relative ranking of these sub-trajectories match that of the complete trajectories they were sampled from. While this assumption is reasonable in certain domains such as Atari and Mujoco, we argue that it can be detrimental in the load balancing domain when the performance is mainly determined by the throughput-based metrics as presented in Section~\ref{subsection:kpis}. It is commonly observed that the network throughput varies significantly through time due to regular daily high and low usage periods. For example, Fig. \ref{fig:trajectory_over_timesteps} shows the variation of the minimum IP throughput among the cells ($T_{min}$) in the span of one week. Clearly, sampling pairs of sub-trajectories from different time intervals would often result in an inconsistency between the assumption from \cite{brown2019extrapolating} and the performance metric. Indeed, we find such inconsistency in $35\%$ of our samples in our experiment. Its adverse effect on the learning capabilities of the reward network is discussed in detail in Section \ref{subsection:reward_extrapolation}.

\begin{figure}
    \centering
        \includegraphics[width=0.85\columnwidth]{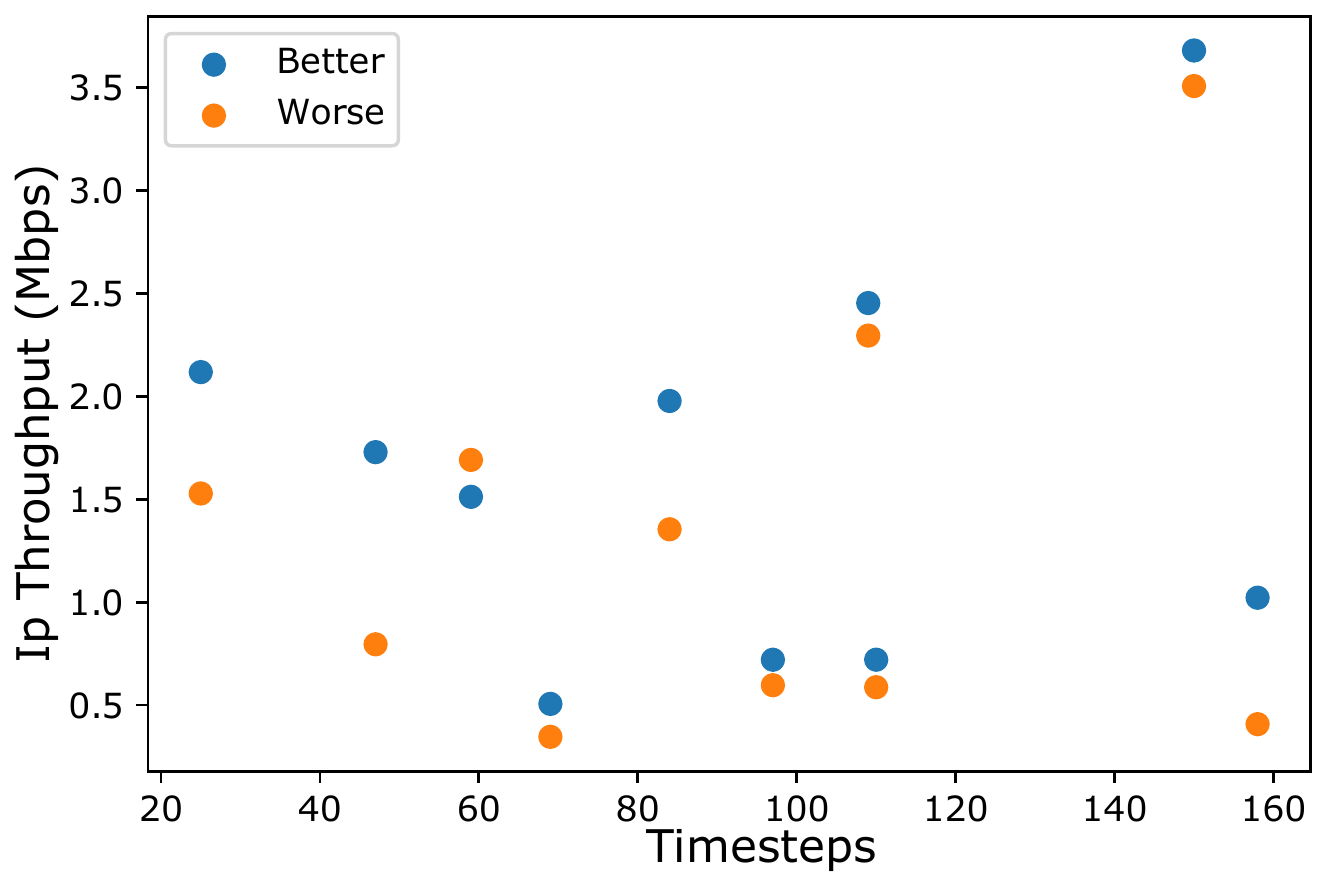}
        \caption{Scatter plot of the minimum IP throughput at different timesteps of two trajectories. While the value across different timesteps varies widely, for a given time, minimum Ip throughput of the state from the better trajectory is consistently (9/10) higher than its worse counterpart. Only 10 such states are shown for visual clarity. } 
        \label{fig:trajectory_over_timestep_zoom}
    \label{fig:indiv_hist-miniptput}
    
\end{figure}

To mitigate the problem, we propose a new sampling procedure: Temporally Consistent Sampling(TCS). TCS introduces two changes to the previously proposed sub-sampling technique. 
Firstly, to account the temporal variation in the network throughput, 
we opt for sampling over random time indexes rather than contiguous blocks. Secondly, as the variation
in the network KPIs across different trajectories at a given time index is relatively lower and fairly consistent with the original relative ranking (Figure \ref{fig:indiv_hist-miniptput}), we mandate temporal consistency,
i.e., given a pair of ranked trajectories, a sub-trajectory should be sampled from the same time indexes for both the trajectories maintaining temporal consistency. Empirically, TCS can reduce the aforementioned inconsistency from $35\%$ to $15\%$, significantly increasing the effectiveness of the training samples.
Our approach is summarized in Algorithm \ref{alg:tcs-trex}.


\begin{algorithm}[t]
\caption{Temporally consistent sampling}
\label{alg:tcs-trex}

\SetKwInOut{Input}{Input}
\SetKwInOut{Output}{Output}

\Input{Ranked demonstrations: $D_E$, Number of samples to generate: $t_{sub}$ 
}
\Begin{
    \For{$i=1$ to $t_{sub}$}{
    Randomly select a pair of trajectories, $t_x$ and $t_y$, from $D_E$ \\
    Define the length of a sample sub-trajectory, $l$. \\
    Randomly select a set of $l$ indexes, $n_l$. \\
    $t_{x}^{samp} \leftarrow t_x[n_l]$, $t_{y}^{samp} \leftarrow t_y[n_l]$ \\
        \eIf{rank($t_x$) $>$ rank($t_y$)}{
            $t_{label} \leftarrow 0$ 
        }{
            $t_{label} \leftarrow 1$ 
        } 
    Append $\{(t_{x}^{samp}, t_{y}^{samp}), t_{label}\}$ to $T_s$. 
    }
    Return $T_s$
}
\Output{Training set, $T_s$}
\end{algorithm}

\section{Experiments and results}
\subsection{Simulation environment}
The experiments are conducted on a  system-level RAN  simulator\cite{slsstart,slsend} consisting of seven eNBs. Each base station consists of three sectors, which in turn comprises of four cells with different carrier frequencies. Six eNBs are arranged in a uniform hexagonal ring with one eNB at the center as shown in Figure \ref{fig:hex7-layout}. The UEs in the system are randomly distributed over the geographical area and are assumed to be points with fixed velocities and random directions drawn from a uniform distribution. We test the extrapolation performance of the reward network in two traffic scenarios and the load balancing performance of the policy network in four scenarios (ID 1-4). These scenarios are determined by different number of UEs, request packet size and interval distributions. 

\subsection{Collecting demonstrations}
For each scenario, we collect a set of $100$ trajectories using a random controller. A random controller randomly samples an action at each hour. In the absence of an external critic to rank the demonstrations, we resort to an ad-hoc ranking function, $R_{f}$, based on a weighted combination of a set of KPIs. We use the reward function from \cite{Li2022clusteringRL} as our ranking function. We would like to emphasise the fact that the use of ranked trajectories provides a significantly low resolution image of the reward landscape than using a reward function to learn a controller in an RL setting which needs access to the reward corresponding to each state it encounters. For a given pair of expert demonstrations $\tau_i$ and $\tau_j$, $\tau_i \prec \tau_j$ when $\sum_{s \in \tau_i} R_{f}(s) < \sum_{s \in \tau_j} R_{f}(s)$. From the set of $100$ trajectories, we select the worst $70\%$ to build our training set and leave the rest to test the extrapolation capabilities of the trained reward network.



\subsection{Training details}
\label{subsection:training_details}
The reward network consists of 3 fully-connected layers with a leaky ReLU activation function \cite{nairRectifiedLinearUnits2010} for the input and the first hidden layer. Other hyperparameters of the training the reward and the policy networks are listed in Table \ref{tab:trex-parameters}. The training time for the reward network took on average roughly 45 hours on an NVIDIA A6000. The policy network is trained using PPO. Both the actor and value network consist of 3 fully-connected layers with $256$ neurons in each layer and use the tanh activation function. All the networks are optimized using Adam \cite{kingma2014adam}. Hyperparameters used for training the policy network are listed in Table \ref{tab:rl-parameters}.

\begin{table}[hbp]
\centering
    \begin{subtable}[b]{0.52\columnwidth}
        \begin{tabular}{|p{2.8cm}|l|}
        \hline
        Learning rate &  $1e-5$\\
        \hline
        Weight decay & $1e-4$ \\ 
        \hline
        No. of epochs &  $1200$ \\ 
        \hline
        No. trajectory pairs & $1000$ \\ 
        \hline
        No. sub-trajectory pairs & $50000$ \\
        \hline
        \end{tabular}
        \caption{Reward learning.}
        \label{tab:trex-parameters}

    \end{subtable} %
    \begin{subtable}[b]{0.46\columnwidth}
        \begin{tabular}{|c|c|}
        \hline
        Learning rate &  $0.0003$\\ 
        \hline
        Weight decay & $1e-4$ \\ 
        \hline
        Total timesteps & $200k$\\
        \hline
        Gamma & $0.97$ \\
        \hline
        Clip range & $0.15$ \\
        \hline
        Batch size & 64\\
        \hline
        \end{tabular}
        \caption{Policy learning.}
        \label{tab:rl-parameters}

    \end{subtable}
\caption{Training hyperparameters for different parts of the learning pipeline.}
\end{table}

\begin{table}[hb]
    \centering
    \begin{tabular}{|p{0.5cm}|p{3.5cm}|p{1.1cm}|p{2cm}|}
    \hline 
    \textbf{ID} & \textbf{Method} & Training & Extrapolation \\
    \hline
    \multirow{2}{*}{1} & T-REX (Original) &  \textbf{0.98} & $0.83$ \\
    &  T-REX + TCS(ours)  & $0.94$  & \textbf{0.94} \\
    \hline
    \hline
    \multirow{2}{*}{4} & T-REX (Original)  & \textbf{0.99} & 0.53 \\
    &  T-REX + TCS(ours)   & 0.98  & \textbf{0.93} \\
    \hline
   \end{tabular}
    \caption{Pearson correlation on the training and extrapolation set for different sub-trajectory sampling techniques. Maintaining temporal consistency between candidate sub-trajectories during sampling from a given pair of trajectories consistently outperforms its counterpart across different scenarios. For reference: a higher value is better.}
    \label{tab:pcorr-table}
\end{table}


\subsection{Results: Reward extrapolation}
\label{subsection:reward_extrapolation}
\begin{figure}[!htp]
    \centering
    \begin{subfigure}[t]{0.49\columnwidth}
        \includegraphics[width=\textwidth]{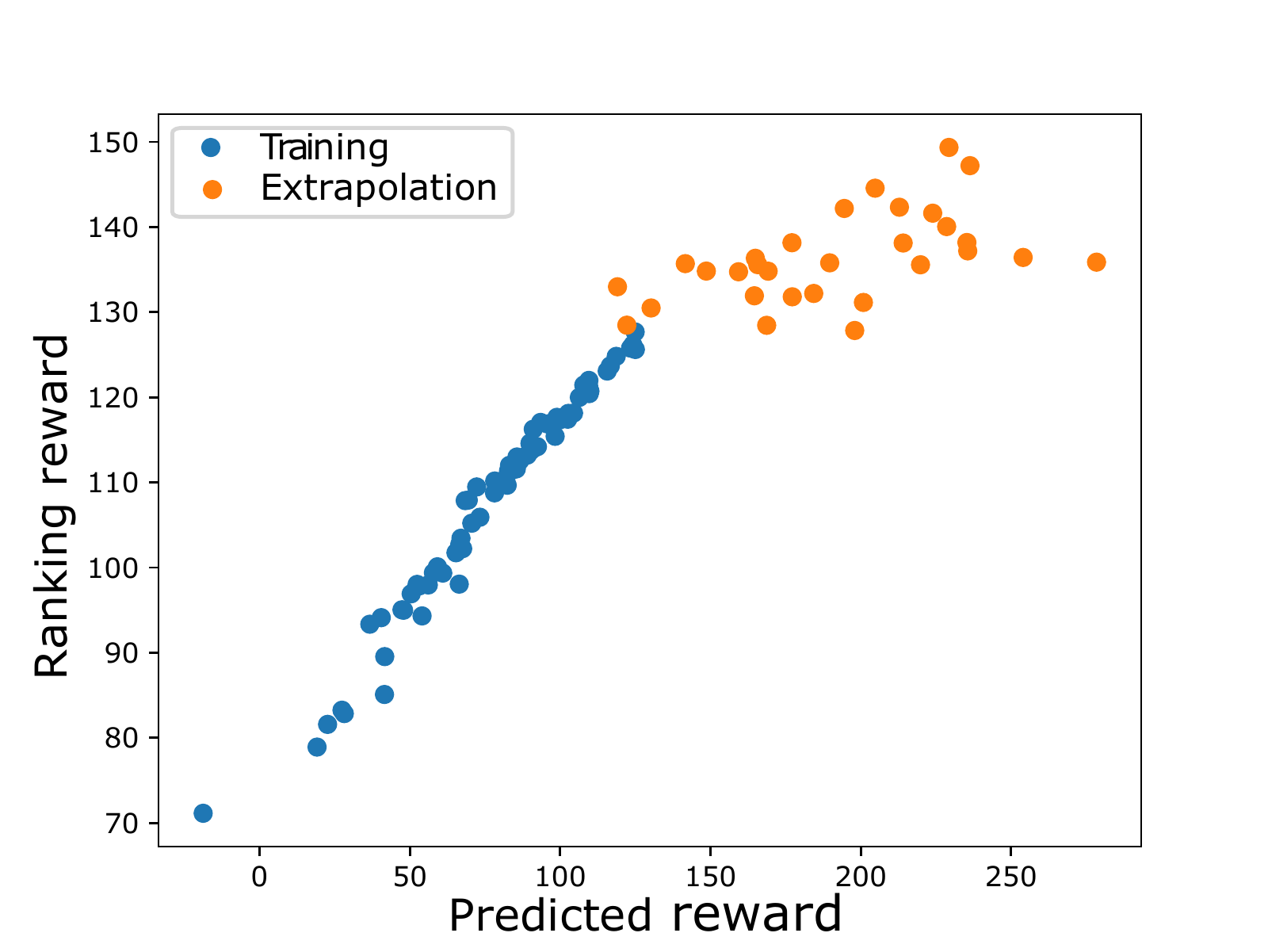}
        \subcaption{Without TCS.}
    \end{subfigure}
    \begin{subfigure}[t]{0.49\columnwidth}
        \includegraphics[width=\textwidth]{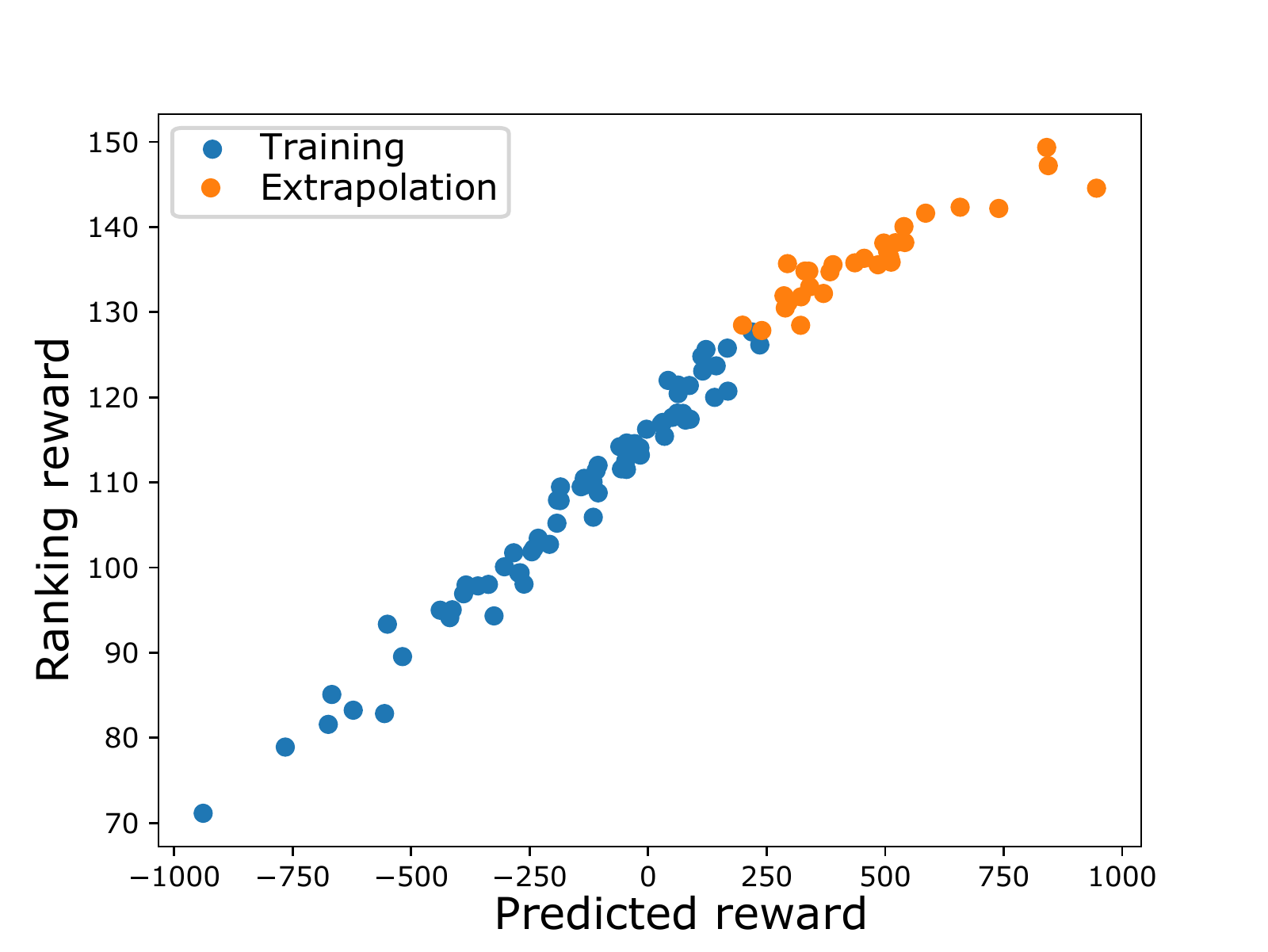}
        \caption{With TCS.}
    \end{subfigure}
    \caption{Scatter plots qualitatively showing the correlation between the predicted and ad-hoc reward for the demonstration trajectories in the training and test (extrapolation) set.}
    \label{fig:pcorr-sector23}
\end{figure}

To test the performance of the algorithm, we use the Pearson correlation coefficient~\cite{wikipedia-contributors-2022}. It calculates the linear relationship between two datasets and outputs a value in the range of $[-1,$ $1]$ which is proportional to their correlation. Table \ref{tab:pcorr-table} shows the correlation values obtained between the ranking reward function and the reward predicted by the trained reward network. From Table \ref{tab:pcorr-table}, we see that while the original method (T-REX) outperforms in the training set, using TCS shows consistent improvement in the extrapolation or test set across different scenarios. This indicates that the model trained from samples generated from the original sampling technique learns a reward function that lacks generalization. A possible explanation for this stems from the empirical observations that the probability of mislabeling a sub-sample pair using TCS drops from $0.35$ to $0.13$, helping the model better capture the latent ranking reward and thus contributing to better extrapolation. A qualitative overview of the extrapolation performance is shown in Figure \ref{fig:pcorr-sector23}.




\subsection{Results: Model performance}
We compare the load balancing performance of the trained policy with four baselines, i.e., demonstrations, fixed rule-based method, adaptive rule-based method and the original TREX method. Demonstrations are the set of trajectories generated from the random controller on which the reward network was trained. Fixed rule-based method is a rule-based method consisting of a set of fixed IULB and MLB parameters as used in \cite{wu2021dataefficientRL}. Adaptive Rule-based method is a rule-based load balancing algorithm \cite{YangAdaptiveRule2012} that dynamically adjusts the IULB and MLB parameters according to the difference in the loads between neighboring cells. We do not include RL based methods in the comparison because these methods usually require access to rewards from individual states as compared to pairwise trajectory ranks utilized by our method. Pairwise ranking on a trajectory level is far more accessible as compared to the former, especially in the real world, and in this work we focus on methods that can be trained and deployed in such conditions.



We evaluate the performance of the  methods on a single sector of the central base station on the three metrics introduced in Section \ref{subsection:kpis}.
Table \ref{tab:kpi-performance-table} reports these metrics averaged over the entire sequence of a given scenario at each hour.

\begin{table}[!h]\small
    \centering
    \def\arraystretch{1}
    \setlength{\tabcolsep}{0.2em}
    \begin{tabular}{|P{1cm}|P{2cm}|P{1.65cm}|P{1.65cm}|P{1.64cm}|}
        \hline
         \textbf{ID} & \textbf{Method} & $\mathbf{T}_{min}$ & $\mathbf{T}_{std}$& $\mathbf{T}_{cc}$\\
         \cline{2-5}
         \hline
         \multirow{4}{*}{1} & Ours & \bm{$2.69\pm0.01$} & \bm{$1.46\pm0.02$} & \bm{$0.00\pm0.00$} \\
         & Fixed rule & $1.60\pm0.01$ & $2.55\pm0.02$ & $0.09\pm0.01$\\
         & Demonstrations & $1.89\pm0.19$ & $2.29\pm0.29$ & $0.05\pm0.07$ \\
         & Adaptive rule & $2.15\pm0.05$ & $2.05\pm0.05$ & $0.03\pm0.01$\\
         & TREX(Original) & $2.51\pm0.02$ & $1.66\pm0.02$ & \bm{$0.00\pm0.00$} \\
        \hline
        \hline
        \multirow{4}{*}{2} & Ours & \bm{$1.97\pm0.02$} & \bm{$1.07\pm0.01$} & \bm{$0.02\pm0.01$}\\
        & Fixed rule & $1.50\pm0.01$ & $1.51\pm0.02$ & $0.03\pm0.01$ \\
        & Demonstrations & $1.45\pm0.13$ & $1.74\pm0.28$ & $0.25\pm0.19$\\
        & Adaptive rule & $1.44\pm0.07$ & $1.89\pm0.10$ & $0.32\pm0.07$\\
        & TREX(Original) & $1.68\pm0.02$ & $1.34\pm0.01$ & \bm{$0.02\pm0.01$} \\
        \hline 
        \hline
        \multirow{4}{*}{3} & Ours & \bm{$1.63\pm0.02$} & \bm{$1.63\pm0.04$} & \bm{$0.20\pm0.02$} \\
        & Fixed rule & $1.47\pm0.02$ & $2.31\pm0.04$ & $0.31\pm0.02$\\
        & Demonstrations & $1.54\pm0.08$ & $2.07\pm0.15$ & $0.30\pm0.10$ \\
        & Adaptive rule & $1.57\pm0.04$ & $2.06\pm0.06$ & $0.28\pm0.03$\\
        & TREX(Original) & $1.62\pm0.02$ & $1.74\pm0.04$ & $0.21\pm0.02$ \\
        \hline
        \hline
        \multirow{4}{*}{4} & Ours & \bm{$2.77\pm0.01$} & \bm{$1.73\pm0.02$} & $0.01\pm0.00$\\
        & Fixed rule & $2.05\pm0.01$ & $2.58\pm0.02$ & \bm{$0.00\pm0.00$}\\
        & Demonstrations & $2.09\pm0.15$ & $2.61\pm0.29$ & $0.02\pm0.03$\\ 
        & Adaptive rule & $2.27\pm0.07$ & $2.41\pm0.12$ & $0.02\pm0.01$\\
        & TREX(Original) & $2.38\pm0.01$ & $2.14\pm0.02$ & $0.01\pm0.00$ \\
        \hline
    \end{tabular}
    \caption{Evaluation results on different traffic scenarios. For $T_{min}$, a higher value translates to a better performance and for $T_{std}$ and $T_{cc}$, lower is better. Our method enjoys an an average improvement of $19.6\%$, $26.7\%$ and $32.3\%$in $T_{min}$, $T_{std}$ and $T_{cc}$ respectively over the next best method.}
    \label{tab:kpi-performance-table}
\end{table}

\begin{figure}[!htbp]
    \centering
    \begin{subfigure}{\columnwidth}
        \centering        

        \includegraphics[width=1\columnwidth]{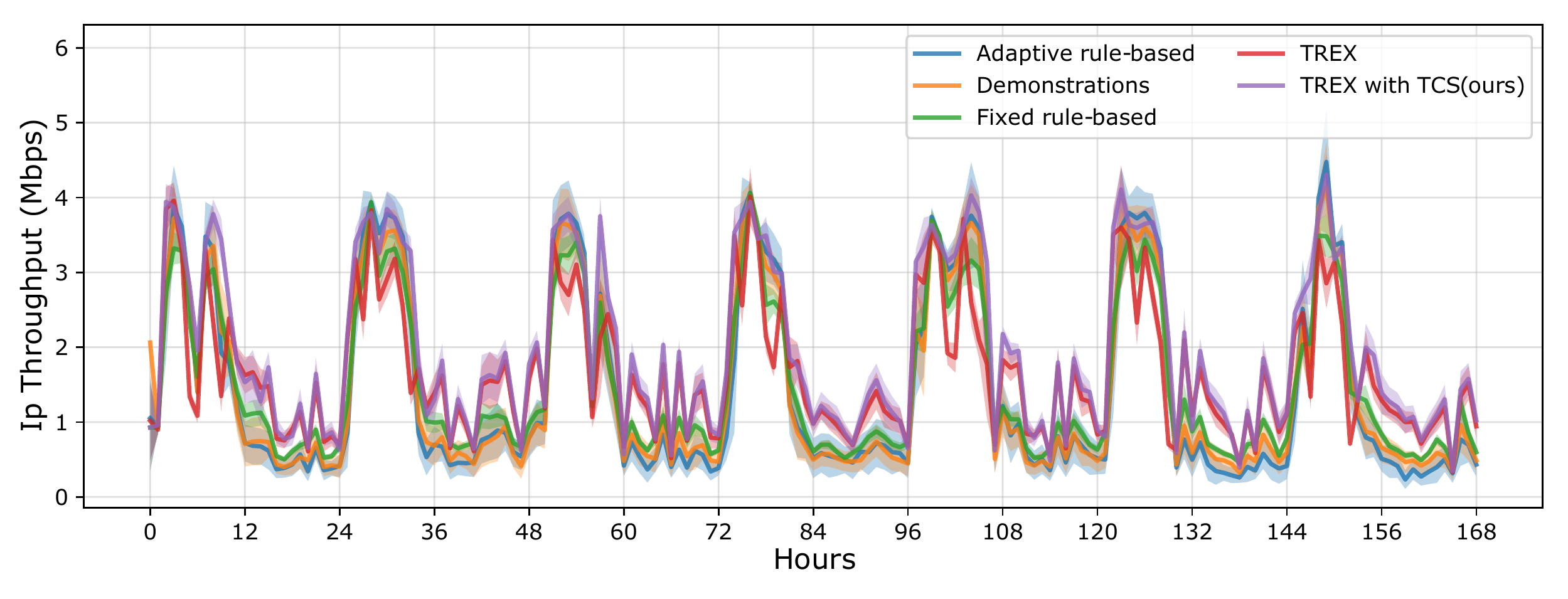}
        \caption{Minimum IP throughput obtained by different methods over the period of one week. From the trends in the minimum IP throughput, the timeline of each week can be divided into low and high traffic intervals. The line plots show that our proposed method better capitalizes on the network resources during moderately low network traffic hours when compared to the other baselines improving the minimum IP throughput in the network.}
        \label{fig:ts-miniptput}
    \end{subfigure}
    \begin{subfigure}{\columnwidth}
    \centering
        \setlength{\belowcaptionskip}{-5pt}
        \includegraphics[width=1\columnwidth]{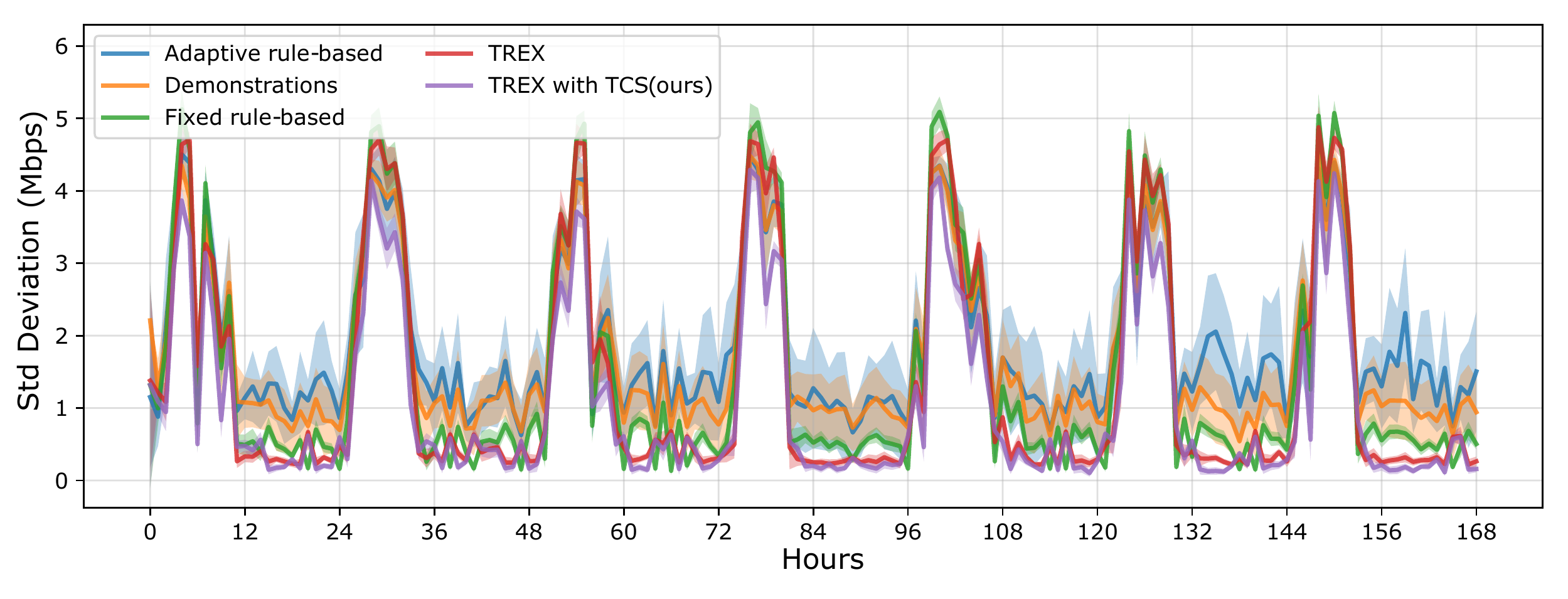}
        \caption{Standard deviation in IP throughput over time. The plot shows that our method consistently maintains a lower standard deviation across the entire week.}
        \label{fig:ts-stdiptput}
    \end{subfigure}
    \caption{Performance of competing methods in different network KPIs.}
    \label{fig:ts-kpi}
\end{figure}

Table \ref{tab:kpi-performance-table}, shows that our proposed method consistently outperforms the other methods across different network scenarios. For $T_{min}$, we achieve on average an improvement of $10.3\%$ over the second best performing method across different scenarios. Figure \ref{fig:ts-miniptput} shows that our method is able to obtain significantly higher $T_{min}$ values at times of moderately high traffic. At times when the network traffic is low, the performance of all the methods is comparable due to the lack of any room for further optimization to wring out more from the existing network infrastructure. Our method also attains a substantial reduction in $T_{std}$. It outperforms its nearest competitor by approximately $14.4\%$ averaged across all the scenarios, and it shows consistent improvement across both periods of high and low traffic as seen from Figure \ref{fig:ts-stdiptput}.


\section{Conclusion and Future work}

With the rapid increase and uneven distribution of communication traffic, communication load balancing has become a pressing problem in maintaining the quality of experience for customers. In this work, we showcase the first attempt to use inverse reinforcement learning for communication load balancing, bypassing the need for an explicitly defined reward signal. We can learn a reward function from a collection of system demonstrations and then utilize that to train a reinforcement learning-based load balancing control policy. Experimental results on different traffic scenarios have showcased the the proposed solution can help significantly improve the system performance. We believe that this work has showcased the effectiveness of inverse reinforcement learning and provides a new direction for future load balancing research.  In the future, we plan to further improve the data efficiency of the proposed solution and investigate the applicability of the proposed solution to other networking optimization problems, e.g., energy saving, network slicing.

\bibliographystyle{IEEEtran}
\bibliography{references}
\end{document}